# Tunable Hyperbolic Phonon Polaritons in a Gradiently-Suspended Van Der Waals α-MoO$_3$

*Zebo Zheng, Fengsheng Sun, Wuchao Huang, Xuexian Chen, Yanlin Ke, Runze Zhan, Huanjun Chen,\* Shaozhi Deng\**

State Key Laboratory of Optoelectronic Materials and Technologies, Guangdong Province Key Laboratory of Display Material and Technology, School of Electronics and Information Technology, Sun Yat-sen University, 510275, China
E-mail: chenhj8@mail.sysu.edu.cn; stsdsz@mail.sysu.edu.cn.

Highly confined and low-loss hyperbolic phonon polaritons (HPhPs) sustained in van der Waals crystals exhibit outstanding capabilities of concentrating long-wave electromagnetic fields deep to the subwavelength region. Precise tuning on the HPhP propagation characteristics remains a great challenge for practical applications such as nanophotonic devices and circuits. Here, we show that by taking advantage of the varying air gaps in a van der Waals α-MoO$_3$ crystal suspended gradiently, it is able to tune the wavelengths and dampings of the HPhPs propagating inside the α-MoO$_3$. The results indicate that the dependences of polariton wavelength on gap distance for HPhPs in lower and upper Reststrahlen bands are opposite to each other. Most interestingly, the tuning range of the polariton wavelengths for HPhPs in the lower band, which exhibit in-plane hyperbolicities, is wider than that for the HPhPs in the upper band of out-of-plane hyperbolicities. A polariton wavelength elongation up to 160% and a reduction of damping rate up to 35% are obtained. These findings can not only provide fundamental insights into manipulation of light by polaritonic crystals at nanoscale, but also open up new opportunities for tunable nanophotonic applications.

Van der Waals (vdW) layered materials and their heterostructures are emerging class of material building blocks for nanophotonic devices that are attracted much recent interests[1-5]. These materials can exhibit novel electronic and optical properties, such as polaritons, hybrid

light−matter modes stemmed from coupling of photons and collective resonances sustained inside the materials [6,7]. Specifically, some of the vdW layered materials such as hexagonal boron nitride (h-BN)[8,9] and molybdenum trioxide (α-MoO$_3$)[10-12], can sustain hyperbolic phonon polaritons (HPhPs), which exhibit extremely high electromagnetic confinements and ultralow optical losses in the infrared to terahertz region[12-14]. These unique properties enable a variety of applications, such as subwavelength focusing[15,16], radiative thermal management [17-19], biochemical sensing[20], Cherenkov radiation[21,22], quantum optics[23], diagnosis of nanostructure defects[24], and energy flow controlling in the nanoscale flatland[25,26].

To better utilize the HPhPs materials for practical nanophotonic components and devices, the external control and active tuning of HPhPs are highly needed, however, which remain challenging. Many attempts have been made to enable the precise tuning and controlling of HPhPs, such as carrier injection, metal intercalation, whereas these paradigms suffer from relatively narrow tunable range and the introduced high optical loss[10,27,28]. Another manner is engineering the surrounding dielectric environment of HPhPs materials. This is because although the hyperbolicity guarantees that HPhPs are confined and guided inside the material, the HPhPs can interact with the local dielectric environment via the evanescent filed at the interfaces[25,29-32]. Specifically, the dispersion relation and propagation loss are strongly dependent on the dielectric function of the surrounding environment[30-32]. This feature allows for flexible tuning and modifying of HPhPs characteristics through controlling the local dielectric environment. For example, it was demonstrated that the wavelength of HPhPs can be tuned by combining h-BN with graphene or phase-changed materials[26,29,32], where the latter exhibit a tunable dielectric function between their metallic and dielectric phases. However, such a tuning approach will usually introduce additional propagation damping induced by the intrinsic dielectric loss of the surrounding materials. An alternative approach demonstrated that the polariton loss from the substrate can be eliminated by isolating the h-BN slab from the substrate underneath, thus the figure of merit of the HPhPs can be significantly increased[33]. Nevertheless, in this previous study, h-BN was approximately treated as an isolated waveguide surrounded by air, thus the interaction between the evanescent field of HPhPs and the substrate was neglected. Such an interaction can also affect the polariton propagation. On the other hand, previous studies have successfully demonstrated that nano-gap between two macro- or nano-structures are usually associated with a variety of enhanced light−matter interaction phenomena, such as highly squeezed electromagnetic fields[34, 35], enhanced energy transfer and conversions[36-39], etc. It is foreseen that stacking heterostructure with nanoscale gap can open up new opportunities for controlling the energy flows at the nanoscale. Therefore, elucidation of how the separation between the polariton crystal and substrate affects the HPhPs propagation behaviors can not only further our insights on the fundamentals of polaritonics, but also help for exploring various potential applications, such as tunable optical waveguide and resonator, as well as near-field heat energy transfer and conversion. However, such an issue still remains unexplored.

In this study, we report the tunable HPhPs in vdW α-MoO$_3$ flake that is gradiently suspended over silicon dioxide substrate with a gradually increased air gap. We explored the effects of gap size on the HPhPs propagation characteristics using infrared nanoimaging technique. Our experiment results, corroborated with theoretical calculations, demonstrate that instead of controlling the substrate dielectric function, both the wavelengths and dissipations of HPhPs in α-MoO$_3$ flake can be tuned by varying the separation between the flake and the substrate. Most interestingly, the evolution of polaritonic behaviors against gap size are opposite for HPhPs in different Reststrahlen bands. Specifically, the wavelengths for HPhPs sustained in lower (upper) Reststrahlen band of α-MoO$_3$ are substantially expanded (shrunk) upon increase of gap size. Simultaneously, the polariton damping rates will be continuously reduced.

The suspended sample was formed by transferring a α-MoO₃ flake onto another one acting as a topographic step, which is supported onto the silicon substrate covered with 300-nm oxide layer (Figure S1, Supporting Information). The transfer was carefully controlled to make one end of the top flake covering the bottom flake, while the other one was laid on the substrate (**Figure 1**a). In such a manner, a gradient air gap of nanometer size was formed between the substrate and top α-MoO₃ flake. Specifically, the gap size $t_{gap}$, i.e., separation between the lower surface of the suspended flake and substrate, decreases gradually from hundreds of nanometers to zero (Figure 1a, lower panel). The entire suspended α-MoO₃ flake remains intact during the fabrication, as confirmed by two-dimensional Raman mapping (Figure S2, Supporting Information). Furthermore, to eliminate strain effect on the lattice structure of the α-MoO₃ flake, thereafter modification of the permittivity and polariton characteristics, we extract and visualize the Raman peak around 820 cm$^{-1}$, which represents the phonon vibration along the [100] direction of MoO₃. As shown in Figure S2b (Supporting Information), the Raman peaks exhibit constant frequencies, indicating that the intrinsic lattice vibration is well preserved in the suspended structure. To experimentally probe the HPhPs of the suspended α-MoO₃, we performed infrared nanoimaging on the sample using a scattering-type near-field optical microscope (s-SNOM)[8,10–12]. As shown in Figure 1a, a sharp metal-coated tip embedded in an atomic force microscope (AFM), with a curvature of $r$ = 20 nm, was illuminated by a focused quantum cascade laser. The operation frequency of the laser ranges from 895 to 1250 cm$^{-1}$. When the sample was scanned beneath the tip, the back-scattered light from the tip was collected using a pseudoheterodyne interferometry and demodulated at third harmonic of the tip vibration frequency (250 kHz). In this manner, both the topography (Figure 1b) and near-field optical intensity mapping (Figure 1c, d) of the sample can be obtained simultaneously.

In a specific s-SNOM measurement, the illumination laser can be strongly compressed by the metallic tip, whereby HPhPs can be launched into the α-MoO₃ flake.[40,41] The polariton waves propagate with a wavelength of $\lambda_p$, which can subsequently interfere with the edge-reflected polariton waves, giving rise to interference fringes with periodicity of $\lambda_p/2$. In addition, the polaritons can also be launched directly by the sample edge due to the scattering of the illumination light[10–12,33]. These edge-launched polaritons can interfere with the tip-launched one and be out-coupled to the detector of s-SNOM, forming different interference fringes with a periodicity of $\lambda_p$. According to the different fringes, we can extract the polariton momentum, $q_p = 2\pi/\lambda_p$, and damping rate, $\gamma$ (or propagation length), from the s-SNOM image recorded at a specific frequency.

Three types of HPhPs can be sustained by the biaxial α-MoO$_3$ crystal, each corresponds to a specific Reststrahlen band where real parts of the permittivity components along the three principle axes, $\varepsilon_x$, $\varepsilon_y$, $\varepsilon_z$, are oppositely signed.[10,12] Specifically, as shown in Figure 1e, two HPhPs modes with positive phase velocities exist in the frequency ranges of 545 to 851 cm$^{-1}$ (Band 1 with Re($\varepsilon_y$) < 0 and Re($\varepsilon_x$), Re($\varepsilon_z$) > 0) and 820 to 972 cm$^{-1}$ (Band 2 with Re($\varepsilon_x$) < 0 and Re($\varepsilon_y$), Re($\varepsilon_z$) > 0). In Band 3 from 958 to 1004 cm$^{-1}$, where Re($\varepsilon_z$) < 0 and Re($\varepsilon_x$), Re($\varepsilon_y$) > 0, the phase velocities of the HPhPs become negative. Figure 1b shows the topography of the Gradiently-suspended α-MoO$_3$ flake, which exhibit a thickness $t_m$ of 195 nm. The $t_{gap}$ at an arbitrary position can be determined by subtracting the thickness of α-MoO$_3$ from the topography height. The gap is thereby shown to increase from the top to bottom in Figure 1b. Figure 1c shows the corresponding s-SNOM image recorded at an excitation frequency of $\omega = 937$ cm$^{-1}$ (Band 2, $\varepsilon_x = -1.374+0.098i$, $\varepsilon_y = 1.377+0.025i$, $\varepsilon_z = 8.315+0.278i$). One can observe that the periodicity of the interference fringes gradually increases with $t_{gap}$, indicating that polariton wavelengths gradually increases against the separation between the α-MoO$_3$ flake and substrate. In contrast, the dependence of fringe periodicity on $t_{gap}$ is distinctly opposite in Band 3. As shown in Figure 1d, which is recorded at $\omega = 990$ cm$^{-1}$ (Band 3, $\varepsilon_x = 0.409+0.046i$, $\varepsilon_y = 1.948+0.019i$, $\varepsilon_z = -1.381+0.120i$), the fringe periodicity gradually reduces upon increasing $t_{gap}$.

To further reveal the dependence of polariton momentum, $q_p$, on the gap size, we analyzed line profiles of the near-field optical signals at different positions of the gradiently-suspended α-MoO$_3$ flake (Figure 1c, d). As shown in **Figure 2**a and b, amplitude profiles along the *x*-axis ([100] crystalline direction) exhibit a series of oscillation maxima, indicating occurrence of polariton interferences. The polariton wavelength $\lambda_p$, and consequently the $q_p$ ($2\pi/\lambda_p$), can then be derived by measuring the separation between adjacent maxima of the profiles. Figure 2c shows that $q_p$ decreases against $t_{gap}$ and gradually saturates upon an excitation frequency of 937 cm$^{-1}$ (Band 2). Conversely, at an excitation frequency of 990 cm$^{-1}$ (Band 3), $q_p$ becomes larger for increased gap sizes, and ultimately saturates for gaps larger than 100 nm (Figure 2d). The $q_p$ at $t_{gap}$ = 200 nm is ~ 1.5 times larger than that of the α-MoO$_3$ portion supported onto the substrate, indicating an enhancement of in-plane optical confinement. In addition, the oscillation strengths of the amplitude profiles are annihilated away from the sample edge, indicating damping of the HPhPs. The damping rate $\gamma$ is defined as Im($q$)/Re($q$), which can be derived by fitting the amplitude profiles. Figure 2e and f show representative amplitude profiles. Dashed lines delegate amplitude envelopes fitted using $y = Ax^{-1/2}\exp(-x/L_0)$, where $L_0 = \dfrac{1}{2\,\text{Im}(q)}$ is the propagation length of HPhPs.[11,29] At $\omega$ = 937 cm$^{-1}$ (Figure 2e), $\gamma$ is reduced to 0.039 and 0.038 for a $t_{gap}$ of 71 nm and 177 nm, respectively. These values are smaller than the supported α-MoO$_3$ can provide, where $\gamma$ is 0.045. An analogous behavior can be observed for a frequency of 990 cm$^{-1}$ (Figure 2f), where the $\gamma$ exhibits a notable reduction up to 35% compared to that of the supported α-MoO$_3$ flake for a gap size of 91 nm. The suppressed polariton damping can be attributed to the reduction of dielectric loss induced by the SiO$_2$ substrate by leveraging the α-MoO$_3$ away from the substrate. These results indicate that not only the polariton wavelength but also the polariton damping can be tuned by varying the $t_{gap}$.

To further corroborate the above experimental findings, we employed an analytical electromagnetic model to calculate the polariton characteristics of the suspended sample. For that reason, the sample is modeled as a multilayer two-dimensional multilayered waveguide as air/α-MoO$_3$/air/SiO$_2$ (Section S1, Supporting Information). The calculated $q_p$ as a function of $t_{gap}$ is shown in Figure 2c and d, manifested as pseudo-color images of the imaginary part of the complex reflectivity $r_p$, Im($r_p$), of the multilayer structure. The calculation results show excellent agreement with the experimental measurements. Moreover, an analytical dispersion relation, $\omega(q_p)$, of HPhPs propagating along $x$-axis in the air/α-MoO$_3$/air/SiO$_2$ multilayer structure can be also determined as (Section S1, Supporting Information)

$$\frac{k_{2z}\varepsilon_s - k_{3z}\varepsilon_a}{k_{2z}\varepsilon_s + k_{3z}\varepsilon_a} = \frac{k_{2z}^2\varepsilon_x^2 - k_{1z}^2\varepsilon_a^2 + 2k_{1z}k_{2z}\varepsilon_x\varepsilon_a \cot(k_{1z}d_1)}{k_{1z}^2\varepsilon_a^2 + k_{2z}^2\varepsilon_x^2} \exp(2k_{2z}t_{gap}) \tag{1}$$

where $k_{1z}$, $k_{2z} = \sqrt{\varepsilon_a k_0^2 - q_p^2}$, and $k_{3z}$ are the $z$-component of photon momenta in α-MoO$_3$, air gap, and SiO$_2$, respectively; $\varepsilon_a$ and $\varepsilon_s$ denote the permittivities of the gap and SiO$_2$, respectively; $d_1$ and $t_{gap}$ are the thickness of α-MoO$_3$ slab and air gap, respectively. The dependences of $q_p$ on $t_{gap}$ calculated by Equation (1) agree well with those obtained from experimental measurements and Im($r_p$) (Figure 2c, d), which therefore corroborate the gap-tuning of the HPhPs in gradiently-suspended α-MoO$_3$. Having establish the validity of the analytical model, we reason that the tunability of the HPhPs by the gap size is associated with the additional factor $e^{2k_{2z}t_{gap}}$, which quantifies the modifications of amplitude and phase of the reflected polariton waves by the interface between α-MoO$_3$ and air gap. This mechanism will be further discussed in the following analyses.

Full-wave simulations using finite element method (FEM) were performed to further unveil the evolution of the optical near-field evolution against $t_{gap}$. The simulation scheme is shown in **Figure 3**a, where a α-MoO$_3$ flake of thickness $t_m$ = 195 nm is suspended over the SiO$_2$ substrate with an air gap of varied sizes. A $z$-polarized electric dipole is placed in proximity above the flake acting as an excitation source. In Figure 3b and c, we plot the cross-sectional (x–z plane including the dipole) snapshots of the z-component of electric field, Re($E_z$). The two images were respectively recorded at excitation frequencies of 937 cm$^{-1}$ and 990 cm$^{-1}$. One can see that for both frequencies, the HPhP waves launched by the dipole source propagate away along the x-axis of α-MoO$_3$ flake. Specifically, for polariton waves in Band 2 (937 cm$^{-1}$), the optical fields are mainly located at the flake surface (Figure 3b), suggesting their surface nature. In contrast, in Band 3 (990 cm$^{-1}$) the polaritons are mainly confined inside the flake (Figure 3c). For both frequencies, there are always evanescent fields confined closely to the upper and lower surfaces of the flake (Figure 3b, c). It is clear that the polariton wavelengths are lengthened (shortened) with increasing $t_{gap}$ at $\omega$ = 937 cm$^{-1}$ ($\omega$ = 990 cm$^{-1}$) (Figure 3b, c). These behaviors are consistent with the experimental and model calculation results (Figure 2c, d). Interestingly, in comparison with the supported α-MoO$_3$, where the electric fields gradually decay into the substrate, presence of the air gap can strongly enhance the vertical confinement of the optical fields. Such confinements can be even down to a few nanometers (Figure 3d, e). The numerical simulated near-field distributions also indicate their steadily decay behaviors when the HPhPs propagate away from the dipole source (Figure 3b, c). The polariton damping rate can thereby be obtained by fitting the line profiles of $E_z$ along the x-axis using the relation $y = Ax^{-1}\exp(iqx)$. As shown in Figure 3f, for $\omega$ = 937 cm$^{-1}$, suppressed polariton damping aganst $t_{gap}$ can be obtained, which is consistent with the experimental results (Figure 2e, f).

To further reveal the underlying mechanisms of the distinct gap-tuning behaviors of HPhPs in the two Reststrahlen bands, one can refer to the Fabry−Pérot resonance condition in the multilayer structure. It is noted that the hyperbolicity of the α-MoO$_3$ flake in the multilayer structure guarantees that the electromagnetic fields are confined and guided by the α-MoO$_3$ slab, despite insertion of the air gap ($\varepsilon_a$ = 1). The polariton waves will experience internal total reflection at the flake interfaces, i.e., the electric fields evanescently penetrate into the substrate and superstrate and then propagate back to the slab (Figure 3b, c). In comparison to HPhPs in α-MoO$_3$ flake supported by the SiO$_2$ substrate, the insertion of an air gap will modify the phase shift of the reflected polaritons waves at the lower interface due to the change of the dielectric difference at the interface. According the Fabry−Pérot resonance condition, the dispersion relation can be written as[8,33]

$$[\text{Re}(q_p) + i\,\text{Im}(q_p)]d_1 = -\psi(l\pi + \Delta\varphi_{up} + \Delta\varphi_{low}), \quad \psi = \frac{\sqrt{\varepsilon_z}}{i\sqrt{\varepsilon_x}} \qquad (2)$$

where $\Delta\varphi_{up}$ = arctan($\varepsilon_0/\varepsilon_x\psi$) and $\Delta\varphi_{low}$ = arctan($\varepsilon_s/\varepsilon_x\psi$) are the phase shifts at the upper and lower surfaces of the suspended flake, respectively. With the presence of the air gap, an effective dielectric function $\varepsilon_{eff}$ is introduced for the lower surface, and therefore $\Delta\varphi_{low}$ = arctan($\varepsilon_{eff}/\varepsilon_x\psi$). Because the evanescent fields penetrate into the air gap and SiO$_2$ with finite depths of ~ $\lambda_p$ (Figure 3d, e), $\varepsilon_{eff}$ can be approximated as $\sqrt{\varepsilon_{eff}} = \sqrt{\varepsilon_{gap}}\frac{t_{gap}}{L} + \frac{(L-t_{gap})}{L}\sqrt{\varepsilon_s}$, with $0 \leq t_{gap} \leq L$. Parameters $\varepsilon_{gap}$ and $\varepsilon_s$ are the permittivities of the gap and SiO$_2$ substrate, respectively[42]; $L$ is the penetration depth of the electromagnetic fields at the lower surface of the suspended α-MoO$_3$ flake. For the spectral range of 820 to 1010 cm$^{-1}$, $\varepsilon_{gap}$ = 1 and 2.73 < $\varepsilon_s$ < 5.98. Consequently, we can obtain $\partial\varepsilon_{eff}/\partial t_{gap} < 0$. According to Equation (2), we get $\partial q_p/\partial t_{gap} \sim -\frac{1}{\text{Re}(\varepsilon_x)}\partial\varepsilon_{eff}/\partial t_{gap}$. Thus, tuning of $\lambda_p$ against $t_{gap}$ is strongly dependent on the sign of Re($\varepsilon_x$). In Band 2 where Re($\varepsilon_x$) < 0, we have $\partial q/\partial t_{gap} < 0$, thus $\lambda_p$ is increased for a larger gap size. In contrast, an opposite behavior occurs in Band 3, where Re($\varepsilon_x$) > 0 and $\lambda_p$ is reduced against $t_{gap}$.

To further demonstrate the HPhPs behaviors of the gradiently-suspended α-MoO$_3$ in a broad spectral range, we explore polariton dispersion relation by performing s-SNOM measurements with varied excitation frequencies in the range of 895 to 1000 cm$^{-1}$. **Figure 4**a–c and Figure 4d–f show the near-field optical images obtained with representative excitation frequencies in Band 2 and Band 3, respectively. Evidently, for HPhPs in Band 2, the evolution of the fringe spacings with the gap size is analogous to that at $\omega = 937$ cm$^{-1}$ (Figure 1c and Figure 3b), while a converse trend mimic that at $\omega = 990$ cm$^{-1}$ (Figure 1d and Figure 3c) can be observed for Band 3. In addition, the dispersion relation can be derived by extracting the polariton momentum $q$ corresponding to monochromatic s-SNOM images obtained at different $\omega$. Figure 4g shows the experimental dispersion relations recoded at three representative gap sizes ($t_{\text{gap}} = 0$, 20 nm, and 100 nm), which agree well with both the analytical results obtained using Equation (2) and Im($r_{\text{p}}$) (colored dash curves and pseudo-colored image). The tunability of polaritonic confinements, quantified by $\Delta\lambda_{\text{p}} = |\lambda_{\text{gap}}-\lambda_{\text{p}}|$, is strongly dependent on $\omega$ in both Band 2 and Band 3. Our results demonstrate that $\Delta\lambda_{\text{p}}$ ranges from ~ 45% to ~ 160% for HPhPs in Band 2. These values suggest a much wider tuning range for the polariton wavelength in comparison with that reported in suspended h-BN. Indeed, the suspended structure elongate the wavelength of Type II PhPs in MoO$_3$, in comparison to that supported on substrate with larger dielectric function, but the PhPs still exhibits ultrahigh confinement with subwavelength scale. The tunability of PhPs wavelength can be useful in many nanophotonic applications. For example, one can achieve resonance in a cavity or antenna with specific size by precisely tuning the wavelength of polaritons[43,44], thus the localized field enhancement can be dynamically controlled. Specifically, this is important to applications such as dynamic control of chemical reaction and energy coupling at the subwavelength scale. Also, the modified dispersion relation of PhPs, showing an increased phase velocity, which results in an acceleration of light, can be achieved in the suspended structure[45]. A relatively narrower tuning range varying from 12% to 36% can be obtained in Band 3. These results also highlight the advantage of HPhPs with in-plane hyperbolicity for tunable nanophotonic devices.

In conclusion, by combining mid-infrared nanoimagings, FEM simulations, and analytical model calculations, we demonstrate that HPhPs in an α-MoO$_3$ flake can be tuned by varying the size of air gap between the flake and substrate. The results show that the polariton wavelengths for HPhPs sustained in lower and upper Reststrahlen bands of the α-MoO$_3$ exhibit converse dependence on the gap size. A tuning range of up to 160 % for the polariton wavelength can be obtained. The polariton damping rates can also be continuously reduced by increasing the size of the gap, whereby a reduction up to 35% of the damping rate can be obtained in the upper Reststrahlen band. Note that only steady-state control of PhPs has been demonstrated in our current study, because the suspended structure was simply fabricated by conventional stacking technique. We only move one small step to the fully control of phonon polaritons for the future applications. To apply the principle to dynamic controllable devices, many methods such as nanoelectromechanical technique can be introduced to actively control the gap[46], thereafter the dynamically controllable phonon polaritons can be achieved. Our findings on one hand help elucidating the mechanisms governing the influences of supporting substrate on the polaritonic behaviors in vdW crystals, and on the other hand provide a promising approach for continuous tuning of the HPhPs in vdW crystals, which therefore can pave the way for design and fabrication of tunable nanophotonic devices.

**Experimental Section**

*Fabrication of the gradiently suspended α-MoO$_3$ 2D flake*: The high-quality α-MoO$_3$ crystals were synthesized by thermal physical deposition method[47]. The gradiently suspended α-MoO$_3$ was prepared using a dry-transfer method. To that end, the thick α-MoO$_3$ flake, acting as a topographic step, was exfoliated from the high-quality α-MoO$_3$ crystal and transfered onto the SiO$_2$ substrate. A piece of poly (bisphenol A carbonate) (PC)/polydimethylsiloxane (PDMS) adhered onto a glass slide was employed to pick up another exfoliated α-MoO$_3$ thin flake. Finally, the PDMS with the picked α-MoO$_3$ flake was stacked onto the topographic step and then released by thermal heating at ~100 ºC.

*Infrared optical nanoimaging* The s-SNOM nano-imaging was conducted using a scattering-type near-field optical microscope (NeaSNOM, Neaspec GmbH). To image the PhPs in real-space, a mid-infrared laser (QCL, Daylight) with tunable wavelengths from 8 μm to 11.17 μm (895 cm$^{-1}$ ~ 1250 cm$^{-1}$) was focused onto the sample with a metal-coated AFM tip (Arrow-IrPt, Nanoworld). During the measurements, the tip was vibrated vertically with a frequency around 250 kHz. The back-scattered light from the tip was demodulated and detected at a third harmonic of the tip vibration frequency.

*2D FFT filtering of the near-field optical images* To extract the tip-launched HPhPs from the near-field optical images, a 2D FFT filtering was employed (Gwyddion). First, the Fourier transform of the image was calculated. Next, a 2D FFT filter is applied to this transform. Within the 2D FFT filter the frequencies that should be removed from spectrum or suppressed to value of neighbouring frequencies were selected by marking appropriate areas in the power spectrum graph. Finally, the inverse transform is applied to obtain a filtered image.

**Supporting Information**
Supporting Information is available from...

**Acknowledgements**
We thank Dr. Yongjun Li from Quantum Design China for his help on the s-SNOM measurement. The authors acknowledge support from the National Natural Science Foundation of China (grant nos. 91963205, 11904420), the National Key Basic Research Program of China (grant nos. 2019YFA0210200, 2019YFA0210203), the Guangdong Natural Science Funds for Distinguished Young Scholars (grant no. 2014A030306017), Guangdong Basic and Applied Basic Research Foundation (grant no. 2019A1515011355, 2020A1515011329). H.C. acknowledge the support from Changjiang Young Scholar Program. Z.Z. acknowledge the project funded by China Postdoctoral Science Foundation (grant no. 2019M663199).

**Conflict of Interest**
The authors declare no conflict of interest.

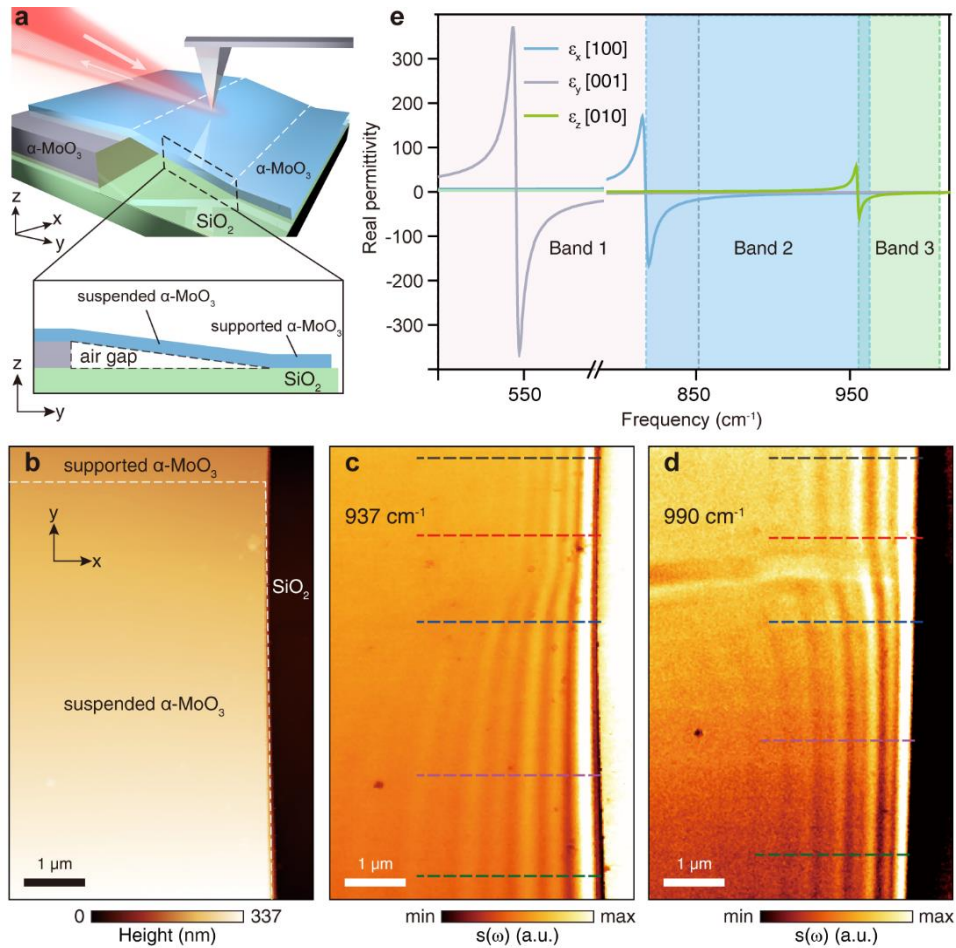

**Figure 1**. Infrared optical nanoimaging of suspended α-MoO₃ flake. **a**) Schematic showing nanoimaging of the suspended α-MoO₃ flake using a scattering-type scanning near-field optical microscope. Lower panel: side view of suspended α-MoO₃ flake with a gradient air gap. **b**) Atomic force microscope image of the suspended α-MoO₃ sample. The white dash lines mark the boundary of the suspended area. **c, d**) Near-field optical images corresponding to the sample in (b). The images are taken at frequencies of 937 cm$^{-1}$ (c) and 990 cm$^{-1}$ (d), respectively. The colored dash lines denote the direction of the line-profile analysis of near-field optical intensities. **e**) Real parts of permittivities along the [100] (*x*-axis), [001] (*y*-axis), and [010] (*z*-axis) crystalline directions of the α-MoO₃. Regions shaded in pink, blue, and green represent Reststrahlen Band 1, Band 2 and Band 3, respectively.

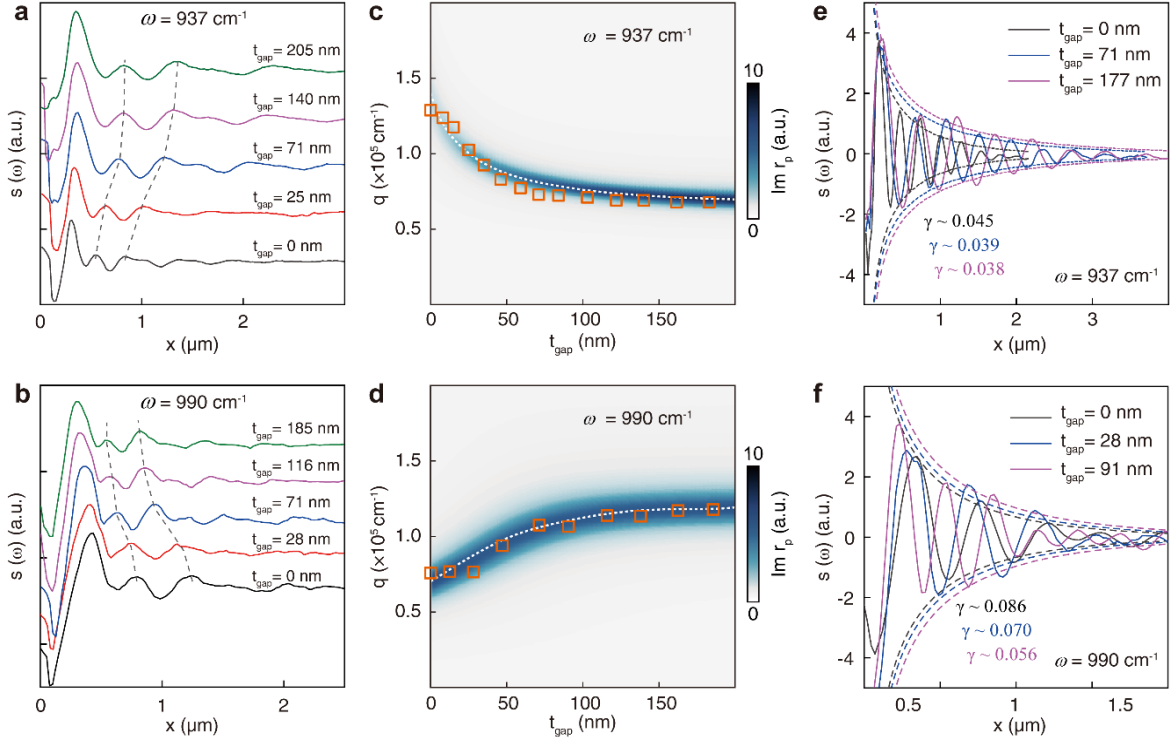

**Figure 2**. Polaritons interferometry in suspended α-MoO₃ flake with varying gap sizes. **a, b**) Near-field amplitude profiles extracted along the dashed lines shown in Figure 1c and d. Grey dashed lines are guide for the eye tracking separations between adjacent maxima. The air gap sizes $t_{gap}$ were extracted from Figure 1b. **c, d**) Dependence of amplitudes of polariton wavevectors on $t_{gap}$. The orange squares indicate experimental measurements. Pseudo-color images demonstrate the $q-t_{gap}$ relation obtained by calculating the imaginary part of the complex reflectivity, Im($r_p$), of the air/α-MoO₃/air/SiO₂ multilayered structure. Dashed white lines denote the explicit $q(t_{gap})$ derived from the analytical electromagnetic model. **e, f**) Line profiles obtained by applying two-dimensional Fast Fourier transform filtering algorithm (2D FFT filtering) on the near-field profiles shown in (a) and (b). Dashed lines represent envelopes of the profiles for three typical $t_{gap}$.

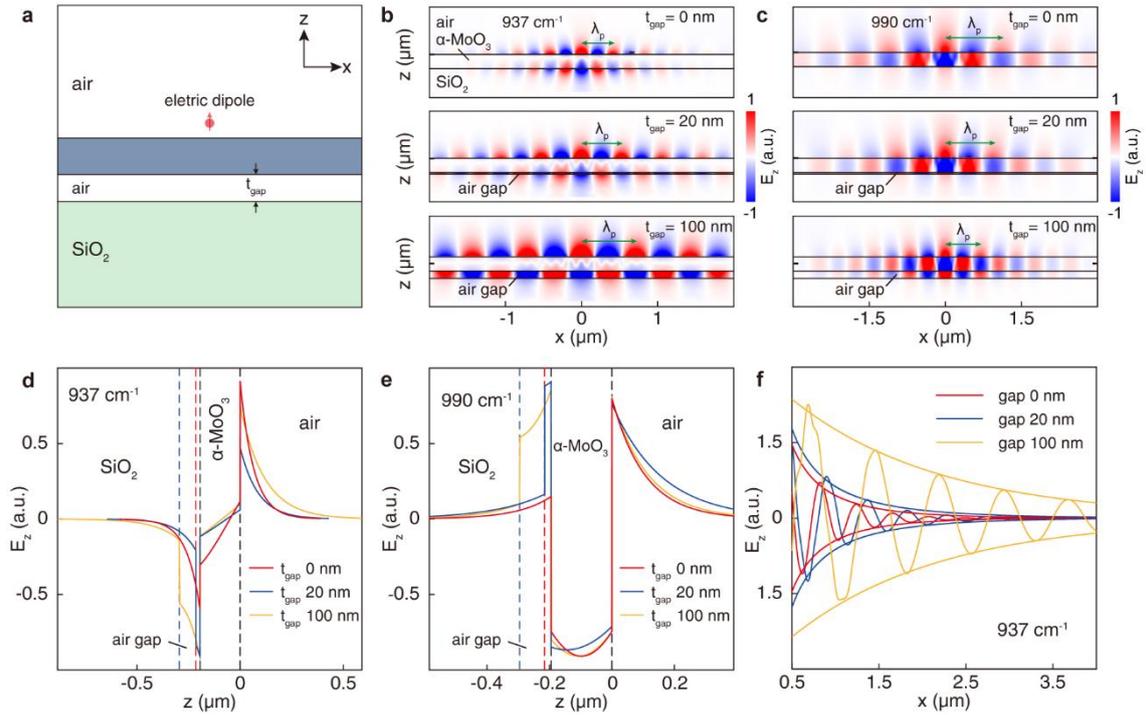

**Figure 3.** Simulations of HPhPs near-fields in gradiently-suspended α-MoO₃ flakes. **a)** Schematic showing the simulation architecture. **b, c)** Simulated Re($E_z$) at frequencies of 937 cm$^{-1}$ (b) and 990 cm$^{-1}$ (c). Three representative gap sizes are considered: 0, 20 nm, and 100 nm. **d, e)** Distributions of $E_z$ along z-axis for the three gap sizes at frequencies of 937 cm$^{-1}$ (b) and 990 cm$^{-1}$ (c). **f)** Line profiles of Re($E_z$) along the upper surface of α-MoO₃ flake extracted from (b).

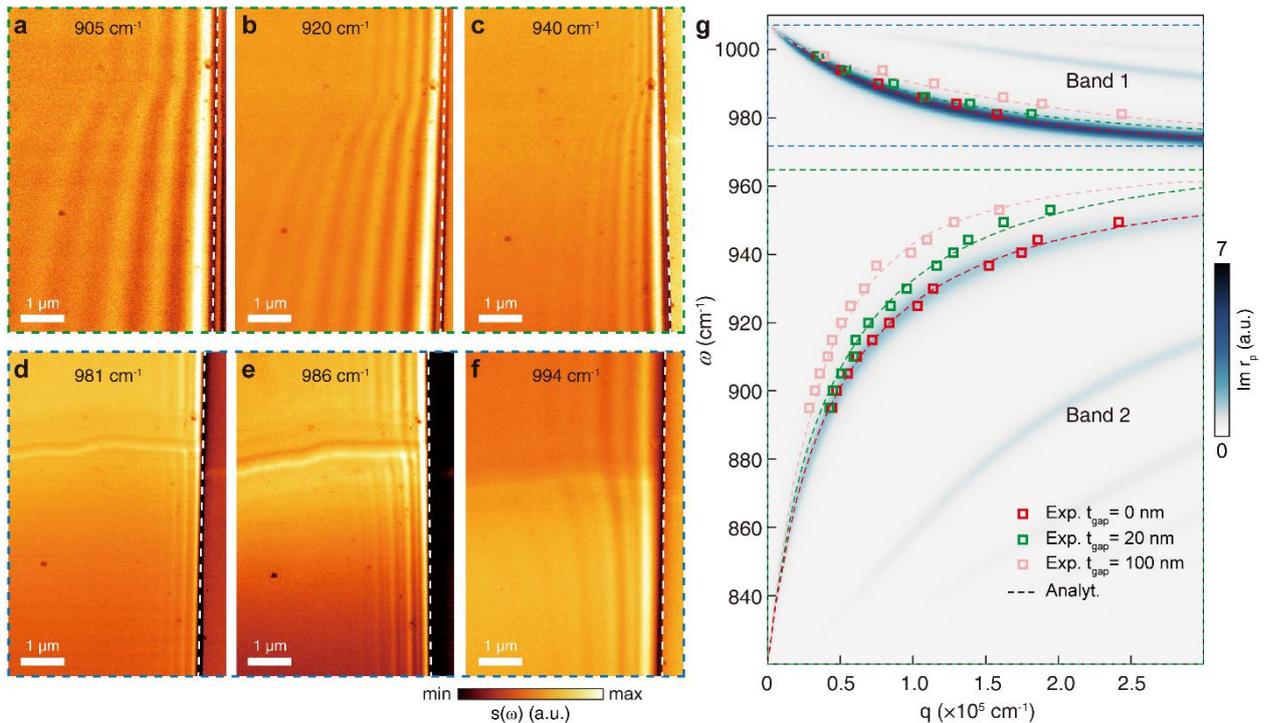

**Figure 4.** HPhPs dispersion relation of a gradiently-suspended α-MoO₃ flake. **a–c)** s-SNOM images recorded with three representative excitation frequencies in Band 3. **d–f)** s-SNOM images recorded with three representative excitation frequencies in Band 2. **g)** Dispersion relation of HPhPs in the gradiently-suspended α-MoO₃ flake. Colored squares indicate the experimental results obtained from s-SNOM images. Dashed colored lines correspond to the analytical dispersion relation. Pseudo-colored images represent the calculated Im$r_p$ of the air/α-MoO$_3$/air/SiO$_2$ multilayered structure.